\documentclass[10pt,conference]{IEEEtran}

\IEEEoverridecommandlockouts

\usepackage{graphicx,times,cite,amsmath, bm, amssymb, epsfig, array, subfigure,algorithm, algorithmic}

\newcommand{\fref}[1]{Fig.~\ref{#1}}
\newcommand{\tabincell}[2]{\begin{tabular}{@{}#1@{}}#2\end{tabular}}

\usepackage{algorithm}
\usepackage{algorithmic}

\usepackage[acronym]{glossaries} 
\newacronym{6G}{6G}{the sixth generation}
\newacronym{IRS}{IRS}{intelligent reflecting surface}
\newacronym{BER}{BER}{bit error rate}
\newacronym{AI}{AI}{artificial intelligence}
\newacronym{SNR}{SNR}{signal to noise ratio}
\newacronym{ML}{ML}{machine learning}
\newacronym{DL}{DL}{deep learning}
\newacronym{DRL}{DRL}{deep reinforce learning}
\newacronym{DDPG}{DDPG}{deep deterministic policy gradient}
\newacronym{IoT}{IoT}{internet of things}
\newacronym{AO}{AO}{alternating optimization}
\newacronym{UAV}{UAV}{unmanned aerial vehicle}
\newacronym{PDF}{PDF}{probability density function}
\newacronym{BS}{BS}{base station}
\begin{document}
\title{\LARGE{IRS-Assisted Lossy Communications Under Correlated Rayleigh Fading: Outage Probability Analysis and Optimization}}
\author{
\IEEEauthorblockN{Guanchang Li\IEEEauthorrefmark{1},
Wensheng Lin\IEEEauthorrefmark{1}, 
Lixin Li\IEEEauthorrefmark{1},
Yixuan He\IEEEauthorrefmark{2},
Fucheng Yang\IEEEauthorrefmark{3},
and Zhu Han\IEEEauthorrefmark{4}}

\IEEEauthorblockA{\IEEEauthorrefmark{1}School of Electronics and Information, Northwestern Polytechnical University,
Xi'an, Shaanxi 710129, China}

\IEEEauthorblockA{\IEEEauthorrefmark{2}Xi’an Modern Control Technology Research Institute, Xi’an 710065, China}

\IEEEauthorblockA{\IEEEauthorrefmark{3}Research institute of Information Fusion, Naval Aviation University, Yantai, China}

\IEEEauthorblockA{\IEEEauthorrefmark{4}University of Houston, Houston, USA}

Email:\ lgc\_\_@mail.nwpu.edu.cn,\{linwest, lilixin\}@nwpu.edu.cn, \\
  1036157229@qq.com,\ fucheng85@sina.com,\ zhan2@uh.edu

%
}

\markboth{}{}
\maketitle

\begin{abstract}
This paper focuses on an \gls{IRS}-assisted lossy communication system with correlated Rayleigh fading.
We analyze the correlated channel model and derive the outage probability of the system.
Then, we design a \gls{DRL} method to optimize the phase shift of \gls{IRS}, in order to maximize the received signal power.
Moreover, this paper presents results of the simulations conducted to evaluate the performance of the \gls{DRL}-based method.
The simulation results indicate that the outage probability of the considered system increases significantly with more correlated channel coefficients. 
Moreover, the performance gap between \gls{DRL} and theoretical limit increases with higher transmit power and/or larger distortion requirement.
\end{abstract}

\glsreset{IRS}
\glsreset{BER}
\glsreset{SNR}
\glsreset{6G}
\glsreset{DRL}
\section{Introduction}
With the rapid development of \gls{AI} technology and \gls{IoT}, the higher data rate and quality of service are in soaring demand \cite{Zuo2023}.
Therefore, many researchers focus on \gls{6G} wireless communication technologies,
as one of the key technologies for \gls{6G}, \gls{IRS} \cite{Jing2024} is an innovative device and has drawn widespread attention in \gls{6G} wireless communication researches and applications.

To date, there are many researches on the optimization problem in different IRS-assisted communication scenarios. 
Wu and Zhang \cite{Wu2019} firstly proposed a method to improve the reflected beam forming of the passive phase shifter at the IRS by joint optimization method for transmission efficiency of communication systems. Chen \emph{et al}. \cite{Chen2024} maximized the energy efficiency for post-disaster communications by optimizing the number of \gls{IRS} reflecting elements.
Yin \emph{et al}. \cite{Yin2023} proposed a mechanism of joint beamforming optimization for multi-IRS based on covariance matrix adaptation evolution strategy.
Jing \emph{et al}. \cite{Jing2023} developed a low-complexity precoding and phase configuration strategy for \gls{IRS}-assisted MIMO millimeter-wave communications.
In \cite{Fang2020}, Fang \emph{et al}. proposed an \gls{AO} algorithm to maximize the energy efficiency by jointly optimizing the phase shift of RIS and the beam forming at the \gls{BS}.

Thanks to the success of \gls{AI} techniques, researchers have introduced machine learning to optimize IRS beyond conventional optimization algorithms.
For example, Feng \emph{et al}. \cite{Feng2022} focused on an \gls{IRS}-assisted non-orthogonal multiple access (NOMA) network system, and utilized \gls{DDPG} algorithm to solve the non-convex \gls{IRS} optimization problem.
In \cite{Zhang2023}, an optimization scheme based on double depth Q network is proposed to increase system capacity for an unmanned aerial vehicle-borne-IRS-assisted communication system.
Li \emph{et al}. proposed a federated deep learning algorithm to obtain the optimal reflecting coefficients of multi-IRS to maximize the total throughput.

Nevertheless, most of the previous work considers the ideal IRS communication environment, i.e., the channel coefficients via different reflecting elements of \gls{IRS} are independent of each other and the communication is lossless.
In reality, the reflecting elements on the same IRS are very close, and hence the channel conditions via close elements are similar, resulting in the correlated coefficients within an \gls{IRS}.

Moreover, the data recovered in destination may contain some errors due to a variety of factors.  
For traditional lossless communications, the error-corrupted data (lossy data) are useless and 	will be discarded, which leads to a significant waste of communication resources.
Nevertheless, in AI-enabled communication systems, such as \gls{IoT} and semantic communication systems, the effective information in the lossy data could be extracted for further processing by AI technologies. 
As long as the lossy data does not exceed a given distortion degree, the lossy data could be still useful for AI-enabled communication systems \cite{Lin2021}.
Thus, lossy communications have considerable application prospects in \gls{6G} communications.

Motivated by the above facts, this paper focuses on the IRS-assisted lossy communications under correlated channel fading.
We analyze the outage probability of the system, and utilize \gls{DRL} method to optimize the IRS reflecting coefficients.
The main contributions of this paper are summarized as follows:
\begin{itemize}
  \item[$\bullet$]
  We propose an IRS-assisted lossy communication system model under correlated channel fading, and obtain the mathematical model of the considered system.
  \item[$\bullet$]
  We perform a mathematical analysis of the system model and then derive the theoretical outage probability under correlated Rayleigh fading in detail.
  \item[$\bullet$]
  We design a \gls{DRL} method to solve IRS optimization problem for the considered system.
  \item[$\bullet$]
  We evaluate the performance of the optimization method and calculate the outage probability through simulations for \gls{IRS}-assisted lossy communications
\end{itemize} 

The rest of this paper is organized as follows.
Section \ref{sec2} introduces the system model and formulates the optimization problem for the considered system.
Section \ref{sec3} derives the outage probability and further designs the DRL-based IRS optimization scheme to minimize the outage probability.
Section \ref{sec4} calculates the outage probability in different distortion requirements and evaluates the training performance of the \gls{DRL}-based method by simulations.
Finally, we conclude this paper in Section \ref{sec5}.

\section{System Model and Problem Formulation}\label{sec2}

\subsection{IRS-Assisted Lossy Communication System}
 
\begin{figure}[!ht]
\centering \includegraphics[width=2.4in]{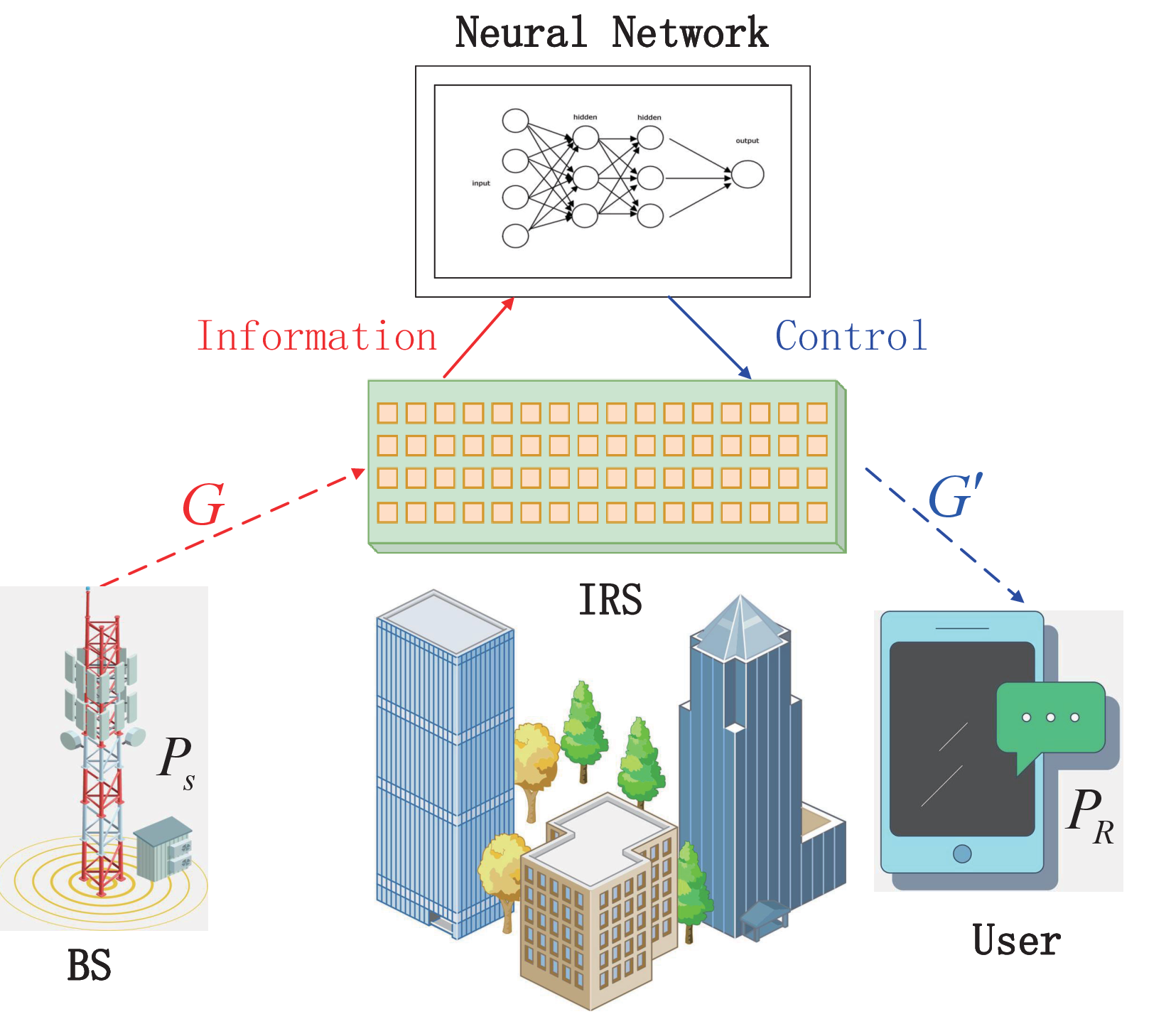}
\caption{IRS-assisted outdoor communication system model.}
\label{fig:sys}
\end{figure}

As illustrated in \fref{fig:sys}, we consider an outdoor \gls{IRS}-assisted lossy communication system.
The signal of direct link is blocked due to the obstacle in the surrounding environment.
The acceptable maximum distortion (i.e., distortion requirement) at the destination is $D$.
To Improve transmission quality, we deploy an IRS with $M$ reflecting elements at an appropriate location to establish an assisted communication link.

Assume that the $P_S$ is the signal transmitting power, $N_0$ is the noise power in the destination and $PL$ is the transmitting path loss of the concatenated BS-IRS-User link.
Let $\bm{G}\in \mathbb{C}^{1\times M}$ and $\bm{G^{'}}\in \mathbb{C}^{1\times M}$ denotes the channel coefficients from \gls{BS} to the IRS, and that from IRS to destination, respectively.
The IRS is divided into $M$ reflecting elements, denoted by $\mathcal{M} = \left \{ 1, 2, . . . , M\right \}$.
Let $\theta_m\in[0,2\pi)$ denotes the phase shift of the $m$-th element, $ m = 1,2,...,M$. 
Then, the IRS reflecting matrix is denoted by $\bm{\Theta} =$ diag$(e^{j\theta_{1}},e^{j\theta_{2}},...e^{j\theta_{M}})\in{\mathbb{C}^{M\times M}}$.
Moreover, the IRS is equipped with a neural network, to optimize the received signal quality of the destination by changing the phase shift of every reflecting element.

\subsection{Correlated Rayleigh Fading Channel} 

\begin{figure}[!ht]
\centering \includegraphics[width=2.6in]{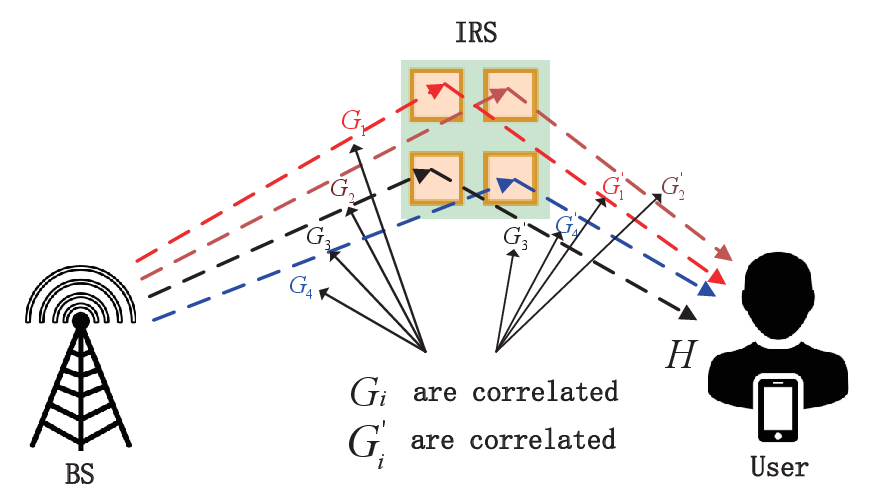}
\caption{Correlated Rayleigh fading channel model.}
\label{fig:channel}
\end{figure}

\fref{fig:channel} illustrates the correlated Rayleigh channel model of the considered system.
As mentioned above, $\bm{G}=(G_1,G_2,...,G_M)$ and $\bm{G^{'}}=(G_1^{'},G_2^{'},...,G_M^{'})$ denote the channel coefficients from \gls{BS} to the IRS, and that from IRS to destination, respectively.
Assuming all links to or from each reflecting elements follow two-dimensional additive white Gaussian noise (AWGN) channel model, the envelope follows a Rayleigh distribution.
In order to describe the channel fading correlation between reflecting elements, we use the model proposed in \cite{Beaulieu2011} to generate $G_i$ and $G_i^{'}$, as
\begin{align}
G_i=&\sigma_i\left(\sqrt{1-\lambda_i^2} X_k+\lambda_i X_0\right) \nonumber \\
    &+j \sigma_i\left(\sqrt{1-\lambda_i^2} Y_i+\lambda_i Y_0\right),
\label{eq:Varience}
\end{align}
\begin{align}
G_i^{'}=&\sigma_i^{'}\left(\sqrt{1-{\lambda_i^{'}}^2} X'_k+\lambda_i^{'} X'_0\right)\nonumber \\
&+j \sigma'_i\left(\sqrt{1-{\lambda_i^{'}}^2} Y'_i+\lambda_i^{'} Y'_0\right),
\label{eq:Variences}
\end{align}
where $j=\sqrt{-1}$, $\lambda_i,\lambda_i'\in{(-1,1)}$ is the correlation factor of $G_i$ and $G_i^{'}$ , and $X_i,Y_i,X'_i,Y'_i\sim\mathcal{N}(0,\frac{1}{2})$ are independent of each other.
Then, $G_i$ and $G_i^{'}$ are two complex Gaussian distribution variable with zero mean respectively, i.e.,\ $G_i\sim\mathcal{N}_c(0,\frac{\sigma_i^{2}}{2}),  G_i^{'}\sim\mathcal{N}_c(0,\frac{{\sigma_i^{'}}^{2}}{2})$.
It can be proved that the correlation coefficient between any $G_i,G_k(i\ne k)$ is $\lambda_i\lambda_k$ and the corresponding power correlation of the resulting Rayleigh fading is equal to $\lambda_i^{2}\lambda_k^{2}$.
When all $\lambda_i^{2}=\lambda_k^{2}=\lambda$, it simplifies to the equal correlation case.
Especially, when all $ \lambda_i = 0 $, it simplifies to the general case of independent channels.

Furthermore, it is easy to know that $\left | G_i \right | $ is a Rayleigh distributed variable with mean-square value $\mathbb{E}(\left | G_i \right |^{2})=\sigma_i^{2}$.
Then, let $\left | \bm{G} \right |=(\left|G_1\right|,\left|G_2\right|,...,\left|G_M\right|)=(r_1,r_2,...,r_M)$, and the \gls{PDF} of $\left | \bm{G} \right |$ is given by
\begin{align}
&f_{\bm{|G|}}\left(r_1, r_2, \ldots, r_M\right) \nonumber \\
= &\quad\int_{t=0}^{\infty} e^{-t} \prod_{k=1}^M \frac{r_k}{\Omega_k^2} \exp \left(-\frac{r_k^2+\sigma_k^2 \lambda_k^2 t}{2 \Omega_k^2}\right) \nonumber \\
&\cdot I_0\left(\frac{r_k \sqrt{t \sigma_k^2 \lambda_k^2}}{\Omega_k^2}\right) d t,
\label{eq:pdf}
\end{align}
where $\Omega_k^2=\sigma_k^2\left(1-\lambda_k^2\right)/2$,
and the analysis above is the same as $G_i^{'}$.
Then, we can derive the compound fading $H$ in destination as
\begin{equation}
H=\sum_{i=0}^{M} G_ie^{j\theta_i}G_i^{'}=\bm{G}\bm{\Theta}\bm{G'}^{T},
\label{eq:fading}
\end{equation}
where the $\bm{G'}^{T}$ denotes the transpose of the matrix $\bm{G'}$.
Therefore, the received power $P_R$ can be denoted by
\begin{equation}
P_R = \frac{P_S \left | H \right |^{2}}{PL}.
\label{eq:power}
\end{equation}
\subsection{Problem Formulation}
In practical communication systems, outage events may occur due to the environment noise and interference.
Consequently, it is necessary to calculate the communication outage probability to evaluate the system reliability under the correlated Rayleigh fading channel.
Based on the outage probability, we can optimize the IRS reflecting coefficients to maximize the system reliability.

Given the channel fading correlation factors $\lambda_i,\lambda_i'$ and variances $\sigma_i,\sigma_i'$, we can generate the correlated channel coefficients $G_i$ and $G_i^{'}$ according to (\ref{eq:Varience}).
From (\ref{eq:fading}) and (\ref{eq:power}), to improve the received power $P_R$, we need to select appropriate phase shift $\theta$ of every reflecting element to maximize $|H|$.
Therefore, the optimization problem can be formulated as
\begin{align}
\mathcal{P}: \max_{\Theta}|H| = \max_{\theta} \sum_{i=0}^{M} G_ie^{j\theta_i}G_i^{'},& \\
\text{s.t.}\quad        \theta_i \in [0,2\pi], \forall i \in{\mathcal{M}}.&
\end{align}

\section{Outage Probability Analysis and Optimization}\label{sec3}

\subsection{Outage Probability Analysis}

For analysis simplicity and without loss of generality, we set $M=4$, and consider a binary Bernoulli source $U$ $\sim$ Bern$(p),p\in[0,0.5]$.

Obviously, if the phase shift of \gls{IRS} is optimal, the received signal components reflected from all reflecting elements should have the same phase.
In this case, the compound signal could achieve the maximum power. 
Hence, we have 
\begin{equation}
\left| H \right|=\sum_{i=0}^{4} \left|G_iG_i^{'}\right| = \sum_{i=0}^{4} \left|G_i\right|\left|G_i'\right|= \left| \bm{G} \right| \left| \bm{G'}^{T} \right|.
\end{equation}
We denote the threshold of the \gls{SNR} in receiver as $\gamma_0$, i.e., the outage event occurs when the \gls{SNR} of received signal is less than $\gamma_0$. 
Then, from (\ref{eq:power}), we can derive the threshold of $\left| {H} \right|$ as
\begin{align}
 H_0 = \sqrt{\frac{\gamma_0 N_0 PL}{P_S}}.
\end{align}
Notice that $\gamma_0$ can be derived from the distortion requirement $D$ by Shannon's source-channel separation theorem \cite{Shannon1993} as
\begin{align}
C(\gamma_0) = R_cR_s(D),
\label{eq:Lossy}
\end{align}
where $C(\gamma)=\frac{1}{2}\log_2(1+\gamma)$ is the Shannon capacity using the Gaussian codebook, $R_c$ is the end-to-end channel coding rate and $R_s(D)$ is the rate-distortion function of the binary Bernoulli source given by
\begin{align}
\begin{array}{cl}
R_s(D)=\left\{
\begin{aligned}
H_b(p)-H_b(D),& 0 \leq D<p.   \\
0, & p \leq D.
\end{aligned} \right. 
\end{array}
\end{align}
where $H_b(x)=-x \log _2 x-(1-x) \log _2(1-x), x \in [0,1]$ is the binary entropy function.
Therefore, the outage probability in IRS-assisted lossy communication system under correlated Rayleigh fading can be calculated as
\begin{equation}
    \begin{aligned}
p_{\text {out }}=\int_A f_{|\bm{G}|}f_{|\bm{G'}|} d r_1 d r_1^{\prime} d r_2 d r_2^{\prime} d r_3 d r_3^{\prime} d r_4 d r_4^{\prime}, \\
A: r_1r_1^{\prime} + r_2r_2^{\prime} + r_3r_3^{\prime}+ r_4r_4^{\prime}<H_0.
    \end{aligned}
\label{eq:pout}
\end{equation}

Next, we start to decompose the integral region $A$.
First, outage events do not occur if $r_1r'_1 \ge H_0$, and hence the outage region $A$ should satisfy the following condition as
\begin{align}
r_1r_1^{\prime}<H_0,
\end{align}
which is equivalent to
\begin{align} 
\left\{\begin{matrix}
0\leq r_1<\infty, \\
0\leq r_1'<\frac{H_0}{r_1}.
\end{matrix}\right.
\end{align}
On this basis, if $r_1r'_1 + r_2r'_2 \ge H_0$, outage events do not occur.
Hence, the outage region $A$ should further satisfy
\begin{align}
r_1r_1^{\prime}+r_2r_2^{\prime}<H_0,
\end{align}
which is equivalent to
\begin{align} 
\left\{\begin{matrix}
0 \leq r_2<\infty, \\
0 \leq r_2<\frac{H_0-r_1r_1^{\prime}}{r_2}.
\end{matrix}\right.
\end{align}
Likewise, we can obtain the conditions of $r_3, r'_3, r_4$ and $r'_4$, and finally derive the conditions for the outage region $A$ as
\begin{align} 
\left\{\begin{matrix}
0 \leq r_i<\infty, \\
0 \leq r_i^{\prime}<\frac{H_0-\sum_{j=1}^{i-1} r_j r_j^{\prime}}{r_i}=\zeta(i).
\end{matrix}\right.
\label{eq:INT}
\end{align}
Thus, (\ref{eq:pout}) can be rewritten by combining with (\ref{eq:INT}) as:
\begin{align}
& p_{out} =\int_0^{\infty} d r_1 \cdots \int_0^{\infty} d r_4 \int_0^{\zeta(1)} d r_1^{\prime} \cdots \int_0^{\zeta(4)} f_{|\bm{G}|}f_{|\bm{G'}|} d r_4^{\prime}.
\end{align}

Notice that  $f_{|\bm{G}|}$ and $f_{|\bm{G'}|}$ contain the modified Bessel function of the first kind with order zero $I_0(\cdot)$, resulting in the difficulty for calculating the integral. 
To solve this problem, we can use a lower-order Taylor expansion term instead of the original function term.
Moreover, the results derived above can be also extended to the \gls{IRS} channel scenario with arbitrary number of reflecting elements.

However, as the number of reflecting elements $M$ grows,  the complexity of $2M$-fold integral in (\ref{eq:INT}) also increases.
Instead of the above methods for calculating the complicated numerical integrals, we can rely on the Monte-Carlo methods \cite{Barbu2020} to obtain the numerical results.

\subsection{IRS Optimization Method Design}
In this section, we develop a \gls{DRL} approach to optimize IRS reflecting coefficients.
As shown in Algorithm \ref{alg:DDPG}, the method has two stages: 1) the global stage for initializing the correlated Rayleigh channel environment as well as the phase control network, and 2) the local stage for reflecting coefficients adjustment strategy.
We define the optimization problem as a Markov decision process, denoted by a transition tuple as $<S_P,A_P,R_P>$, 
where $S_P$ is the state space of the each reflecting element, $A_P$ is the action space of the phase shift, and $R_P$ is the reward.

\begin{algorithm}
\caption{\gls{DDPG}-based IRS Optimization}
\label{alg:DDPG}
\begin{algorithmic}[1]
\REQUIRE {correlation coefficient array $\lambda$ and variance array $\sigma$} 

\textbf{Initialize} correlated channel environment $E$, memory capacity and DDPG agent;
\FOR {sample $m=0$ \TO $N_s-1$} 
\STATE {Generate a sample of channel fading $\bm{G}_m$ and $\bm{G'}_m$ by formula \eqref{eq:fading};}
\FOR {episode $e=0$ \TO $N_e-1$} 
	\IF {replay memory is full}
	\STATE {Set phase shift to $(0^{\circ},0^{\circ},0^{\circ},0^{\circ})$}
	\ELSE 
	\STATE {Set phase shift to a random state;}
	\ENDIF
	\FOR {step $t=0$ \TO $N_t-1$} 
	\STATE {Select IRS phase shift action $a^t$ based on the state $s^t$ by exploratory 		variance strategy $\epsilon$;}
	\STATE {DDPG agent execute the action, get the reward $r^{t}_s$ and next state 	$s^{t+1}$;}
	\STATE {Store the transition ($s^t$,$a^t$,$r^{t}_s$,$s^{t+1}$) in replay memory;}
	\ENDFOR

	\IF {replay memory is full}
\STATE{Decrease $\epsilon$ to reduce the randomness of action selection;}
\STATE{Sample random batch from memory pool;}
\STATE{Perform negative gradient descent to update the evaluation network weights;}
\STATE{Replace target network with current evaluation network every $N_r$ times weights updates.}
\ENDIF
\ENDFOR
\ENDFOR
\ENSURE {The weights of target network, the algorithm runtime and the reward array of each episode.}
\end{algorithmic}
\end{algorithm}

Let $s^{t}_P = {s^t} \in{<S_P>},\forall t$ be the current IRS reflecting coefficient at the time index $t$, with $s^t= (\theta_{1},\theta_{2},...,\theta_{M})\in{\mathbb{R}^{1\times M},\theta_i\in(0,2\pi)}$.
Let $a^{t}_p = \{\Delta\theta^{t}\} \in{<A_P>},\forall t$ be the action for \gls{IRS} phase shift control, where $\Delta\theta^{t} \in{\mathbb{R}^{1\times M}}$ is defined as $(\Delta\theta_1,\Delta\theta_2,...,\Delta\theta_M),\Delta\theta_i\in(-\Delta_m,\Delta_m)$, and $\Delta_m \in [0,90^{\circ})$ can be presetting.
With the aim of maximizing received power $P_R$ and combining with (\ref{eq:power}), we let $r^{t}_s = \left | H \right |^{t}_s \in{<R_P>}, \forall t$ denote the reward after action execution on state $s$.

At the global stage, we first initialize the correlated channel environment, and then build up the \gls{DDPG} neural network to control the phase shift of each reflecting elements.
The \gls{DDPG} neural network consists of two pairs of Actor-Critic networks, including a target network and an evaluation network.
It is noticed that the paired Actor and Critic networks share the hidden layers to reduce the number of model parameters for the purpose of sharing to share some feature extraction capabilities.
Firstly, the environment generates a sample of channel fading $\bm{G}$ according to \eqref{eq:fading} to start optimization.
Then, the evaluation Actor network accepts \gls{IRS} state condition $s^t$, correlation coefficient vector and variance vector to choose the action $\tilde{a}^t$. Subsequently, it adds independent Gaussian noise to each action dimension of $\tilde{a}^t$ to obtain the final action $a^t$, which provides the randomness of action exploration.
After executing $a^t$ and getting the reward $r^{t}_s$ from the environment, the transition tuple will be stored in the memory pool if the pool is not full.

When the memory pool is full, the system starts to update the evaluation network weights,  and decrease the action selection noise gradually at the local stage.
In order to maximize the reward, the agent first samples a random batch and performs gradient descent to update the evaluation network weights based on the target network.
After $N_r$ times weights updates, the agent will update the target network weights with current evaluation network weights. 
At last, the whole system will repeat the above process until the reward convergences.
\section{Performance Analysis}\label{sec4}

In this section, the setting of the system parameters is listed in Table \ref{table:system_parameters}.
We assume the channel coefficients between the BS-IRS link and the IRS-User link are independent, while the channel coefficients within the same link are correlated. Consequently, we need two sets of correlation factors $\bm{\lambda}_1$ and $\bm{\lambda}_2$ and variance vector $\bm{\sigma}$ to describe the channel model.
Moreover, the hyperparameters of \gls{DRL} method is present in Table \ref{table:DRL_ hyperparameter} and all neural networks in the algorithm are implemented based on the Pytorch framework.

\begin{table}[!ht]
\renewcommand{\arraystretch}{1.3}
\caption{System Parameters}
\label{table:system_parameters}
\centering
\begin{tabular}{| c | c | c | c |}
\hline
 Parameter &  Value & Parameter &  Value  \\
\hline
 $PL$ & $40$ dB & $M$ & $4$\\
\hline
 $\bm{D}$ & $(0,0.1,0.2,0.4)$ &  $p$ & $0.5$\\
\hline
 $\bm{\sigma}$ & $(1.0, 1.0, 1.0, 1.0)$& $N_0$ & $-40$ dBm \\
\hline
$\bm{\lambda}_1$ & $(0.95, 0.9, 0.9, 0.85)$ & $\gamma_0$ & $10$ dB \\
\hline
$\bm{\lambda}_2$ & $(0.9, 0.95, 0.85, 0.9)$ & $R_c$ & $1$\\
\hline

\end{tabular}
\end{table}

\begin{table}[!ht]
\renewcommand{\arraystretch}{1.3}
\caption{\gls{DRL} Method Hyperparameters}
\label{table:DRL_ hyperparameter}
\centering
\begin{tabular}{| c | c | c | c |}
\hline
 Hyperparameter &  Value & Hyperparameter & Value \\
\hline
\tabincell{c}{Learning rate of  Actor network } & $0.001$ &  $N_e$ & $300$ \\
\hline
\tabincell{c} {Learning rate of   Critic network} & $0.003$& $N_t$ & $200$ \\
\hline
Replay memory size & $10000$ & $\Delta_m$ & $15^{\circ}$ \\ 
\hline
Batch size & $128$ & Optimizer & Adam \\
\hline
Activation function & tanh() & $N_r$ & $90$ \\
\hline
Exploration variance $\epsilon$ & $3$ & $N_s$ & $10^{5}$\\
\hline
\end{tabular}
\end{table}

 \begin{figure*}
\centering
\subfigure[The memory pool is not full, when episode = 45.]{\label{fig:MemoryStage}
\includegraphics[width=0.3\linewidth]{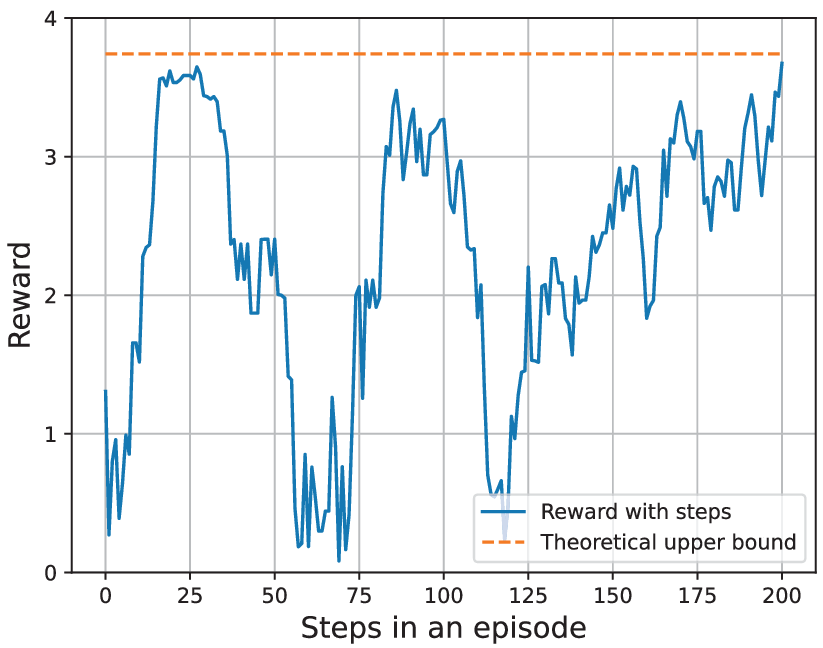}}
\hspace{0.01\linewidth}
\subfigure[The action exploration and stage, when episode = 76.]{\label{fig:TrainStage}
\includegraphics[width=0.3\linewidth]{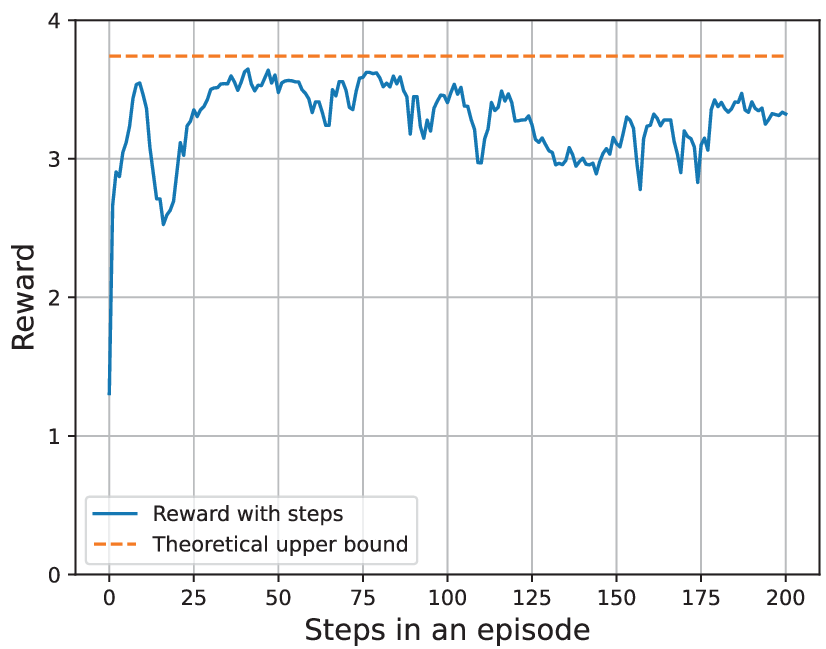}}
\vfill
\subfigure[The local oscillation due to the local optimal, when episode = 94.]{\label{fig:WaveStage}
\includegraphics[width=0.3\linewidth]{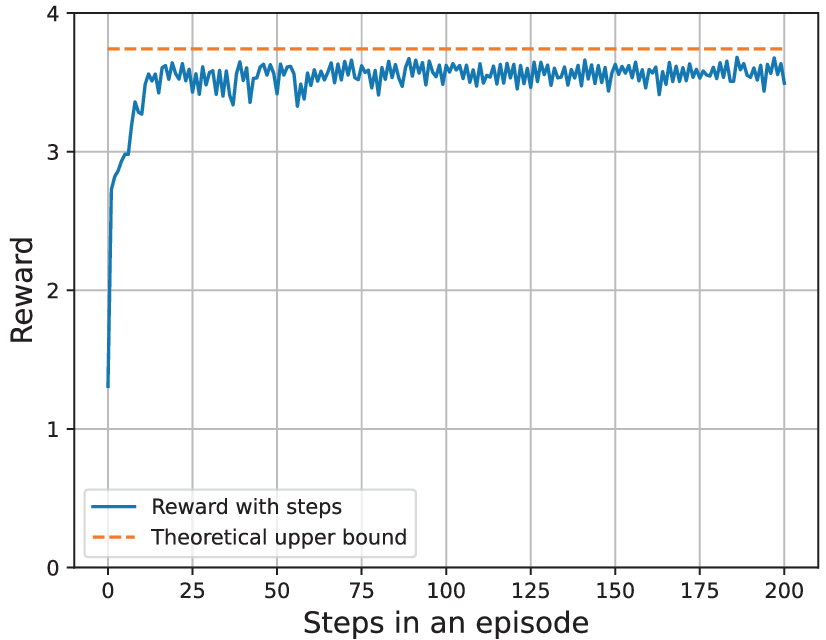}}
\hspace{0.01\linewidth}
\subfigure[The neural network has trained fully and the reward has converged, when episode = 118.]{\label{fig:ConvergeStage}
\includegraphics[width=0.3\linewidth]{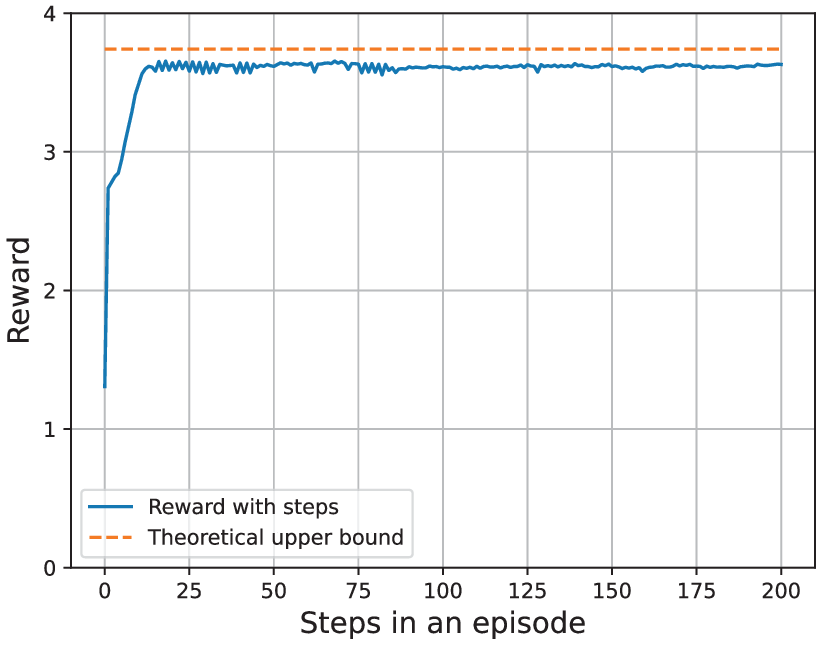}}
\caption{Performance of the proposed algorithm for diverse episodes.}
\label{fig:DRL}
\end{figure*}

%
%
%
%
%
%

\subsection{DRL-Based Method Analysis}
\fref{fig:DRL} shows the value of the reward function $|H|$,  with steps in different episodes for one sample of $\bm{G}$ and $\bm{G'}$. 
\fref{fig:MemoryStage} demonstrates the initial stage when the memory pool is not full.
The initial state is randomly generated within the valid range, and the action is also randomly selected to provide a wider coverage of state-action space for exploration.
Hence, the curve shows a jagged shape with large changes in amplitude.
In \fref{fig:TrainStage}, the memory pool is full and the weights of evaluation network have been updated for several times.
The trend of the curve becomes clear and slowly approaches the theoretical limit.
However, the curve is not smooth enough and there are still many significantly jagged parts. 
This is because the action selection noise $\epsilon$ is still large, leading to the agent choosing actions regardless of whether they are good or bad.  
As the episode of training increases, the curve approaches the theoretical curve smoothly and oscillates at some locations in \fref{fig:WaveStage}, which means the agent can choose a better solution for most states.
However, for a small number of states, it will fall into local optimal solutions.
This is due to the periodicity of the signal phase, which makes the state $s^t$ and the next state $s^{t+1}$ repeat between several values.
Due to action selection noise, this phenomenon will be alleviated after a period of time.
In \fref{fig:ConvergeStage}, the agent will finally converge after iterative episodes.
The curve shows that the agent can stabilize the system reward close to the theoretical optimal value through 15 steps of interaction with environment.

\begin{figure}[!ht]
\centering \includegraphics[width=2.2in]{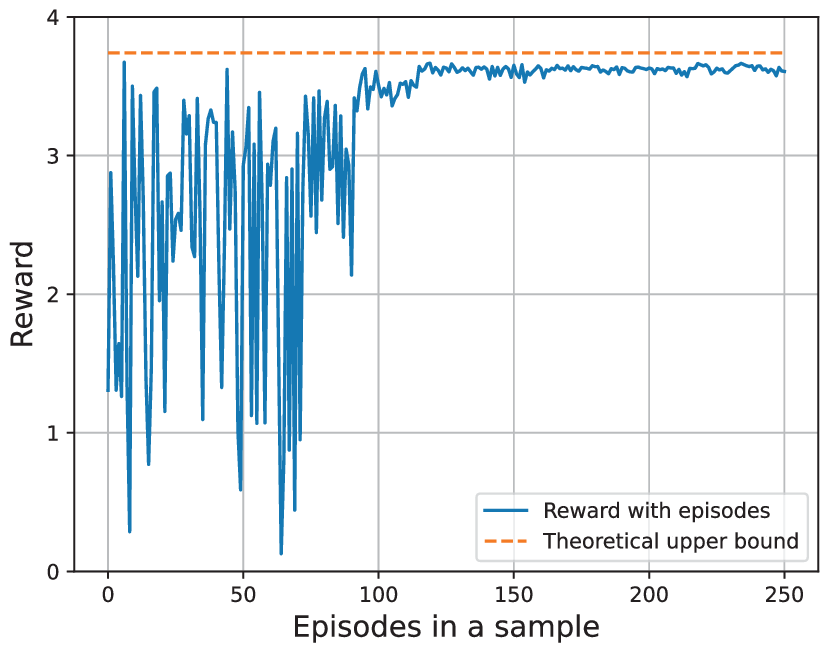}
\caption{Reward with episodes in a sample}
\label{fig:RWE}
\end{figure}
 
\fref{fig:RWE} depicts the last step reward as the episode increases in a sample.
Obviously, the last step reward of episode s unstable during the stage for filling  the memory pool.
As the network training proceeds, the reward gradually approaches the theoretical limit and finally stabilizes near the theoretical limits from the 120 th episode. 

\subsection{Outage Probability Analysis}

\begin{figure}[!ht]
\centering \includegraphics[width=2.7in]{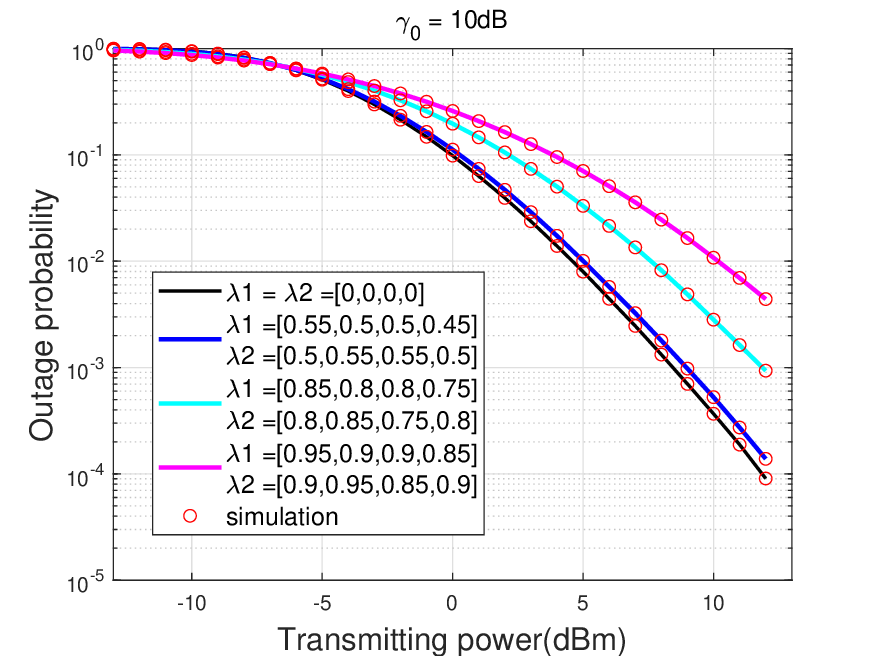}
\caption{Theoretical outage probability with $D=0$.}
\label{fig:Outage}
\end{figure}
\begin{figure}[!ht]
\centering \includegraphics[width=2.7in]{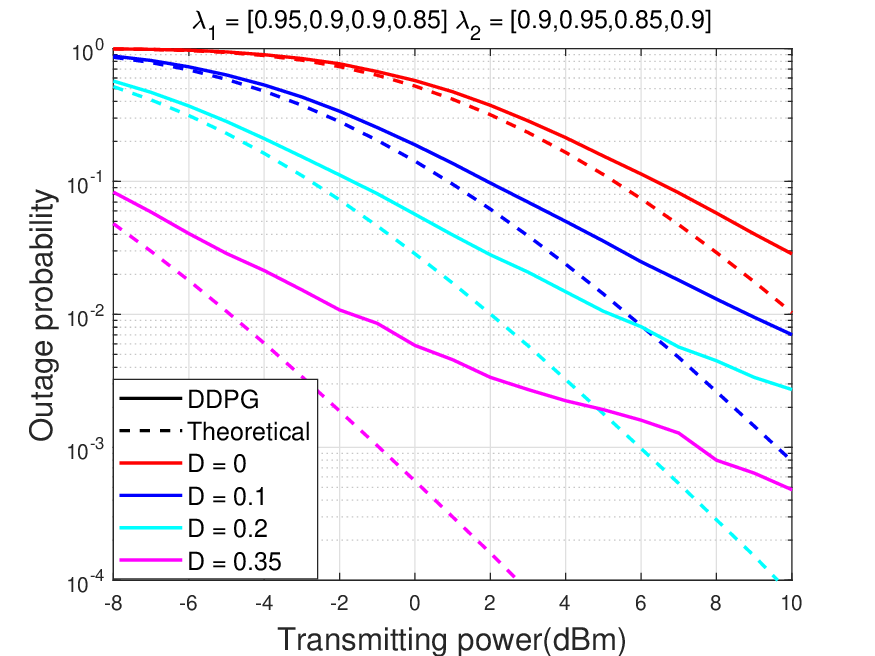}
\caption{The outage probability optimized by DDPG in different distortions.}
\label{fig:Distortion}
\end{figure}

As illustrated in \fref{fig:Outage}, we depict the theoretical and Monte-Carlo simulation results of the outage probability with $D=0$ and $\gamma_0 = 10$ dB for different channel correlations. 
Clearly, the curves obtained by numerical integrals and Monte-Carlo methods perfectly coincide with each other, which verifies the correctness of the theoretical analysis.
Moreover, comparing to independent channel fading, i.e., $\bm{\lambda}_1 = \bm{\lambda}_2 = [0,0,0,0]$, the communication outage probability under correlated channel fading becomes larger as the correlation increases.
This is because the increasing of fading correlation means that multiple signals may experience large fading simultaneously.
This leads to greater signal energy loss, which results in the increment of outage probability.

In \fref{fig:Distortion}, we also present the outage probability based on the data from the \gls{DRL} optimization designed above and compare it with theoretical curves under different distortion requirement $D$.
The performance gap between \gls{DDPG} and theoretical limit becomes larger as the transmit power and/or the distortion requirement increases. 
The reason is that the \gls{IRS} can only control the phase shift, and hence the power loss is proportional to the transmit power, resulting in larger performance loss for higher $P_S$.

\section{Conclusion}\label{sec5}
This paper has investigated an IRS-assisted lossy communication system with correlated Rayleigh channel fading.
In order to further analyze the communication performance of the system, we obtain the mathematical model of the correlated Rayleigh channel and then derive the outage probability of the IRS-assisted lossy communication system based on Shannon's source-channel separation theorem.
After that, we design a \gls{DDPG}-based method to optimize the phase shift of \gls{IRS} to maximize the power of received signal.
Finally, we evaluate the performance of optimization method by simulations,
and calculate the outage probability based on Monte-Carlo methods to and compare it with results obtained from numerical integrals for various distortion requirements.

\section*{Acknowledgement}
The authors would like to thank Prof. Tad Matsumoto of JAIST for the suggestions in English revision.

\bibliographystyle{IEEEtran}
\bibliography{myreference}

\end{document}